\documentclass[10pt,a4paper]{article}
\usepackage{amsmath}
\usepackage{mathptmx}
\usepackage{amsfonts}
\usepackage{amssymb}
\RequirePackage{flushend}
\RequirePackage[numbers,sort&compress]{natbib}
\RequirePackage[colorlinks,citecolor=darkred,urlcolor=magenta,linkcolor=darkblue,bookmarks=false]{hyperref}
\usepackage[pdftex]{graphicx}
\usepackage{subfig}
\usepackage{color}
\usepackage{mathptmx}
\usepackage[misc]{ifsym}
\usepackage{adjustbox}
\usepackage{floatrow}
\usepackage{multirow}
\usepackage{makecell}

\definecolor{darkblue}{rgb}{0.0, 0.0, 0.62}
\definecolor{deepmagenta}{rgb}{0.8, 0.0, 0.7}
\definecolor{darkred}{rgb}{0.55, 0.0, 0.0}

\RequirePackage{./mymacro}
 \graphicspath{{./graphs/}}

\usepackage[left=2cm,right=2cm,top=2cm,bottom=2cm]{geometry}
\author{Srijita Sinha\footnote{\hspace{0.05cm}\Letter \hspace{0.1cm}\href{mailto:ss13ip012@iiserkol.ac.in}{ss13ip012@iiserkol.ac.in}}\hspace{0.15cm} and Narayan Banerjee\footnote{\hspace{0.05cm}\Letter \hspace{0.04cm} \href{mailto:narayan@iiserkol.ac.in}{narayan@iiserkol.ac.in}} \\ IISER Kolkata, Mohanpur Campus, Mohanpur, Nadia 741246, India}

\title{Perturbations in a scalar field model with virtues of \lcdm}
\begin{document}
\date{}
\maketitle


\begin{abstract}
In the era of precision cosmology, the cosmological constant $\Lambda$ gives quite an accurate description of the evolution of the Universe, but it is still plagued with the fine-tuning problem and the cosmic coincidence problem. In this work, we investigate the perturbations in a scalar field model that drives the recent acceleration in a similar fashion that the cosmological constant does and has the dark energy (DE) density comparable to the dark matter (DM) energy density at the recent epoch starting from arbitrary initial conditions. The perturbations show that this model, though it keeps the virtues of a \lcdm model, has a distinctive qualitative feature, particularly it reduces the amplitude of the matter power spectrum on a scale of $8 h^{-1}\, \mpc$, $\se$ at the present epoch. 
\end{abstract}

\vspace{0.5cm}


\vspace{0.5cm}
\vspace{0.5 cm}


\section{Introduction}

The recent cosmological observations using various independent observational data like the Type Ia Supernovae (SNe Ia) measurements~\cite{riess1998aj, schmidt1998aj,perlmutter1999aj,scolnic2018aj}, cosmic microwave background (CMB)~\cite{eisenstein1998aj, planck2015cp, planck2018cp}, Particle Data Group~\cite{partphys2018prd}, large scale structure (LSS)~\cite{reid2010mnras,descol2019prl,alam2017sdss3,alam2020sdss4} show that the Universe is expanding with an acceleration for the past several Giga years. An exotic component called `dark energy' (DE), in the Universe, can help overcome the attractive nature of gravity and make matter move away from each other at a faster rate. To drive the acceleration of the Universe, the pressure ($p$) of the DE must be sufficiently negative, making its ratio with the energy density ($\rho$) at least less than $-\frac{1}{3}$ ($p/\rho = w < -1/3$). A non-zero cosmological constant, $\Lambda$, is undoubtedly the preferred one~\cite{paddy2003pr,copeland2006ijmpd,frieman2008araa,amendola2010prl,mehrabi2018prd,planck2018cp}. A scalar field with a potential, introduced by Peebles \& Ratra~\cite{peebles1988apjl} and Ratra \& Peebles~\cite{ratra1988prd}, is also a popular choice. A lot of work followed from there for various purpose. Some examples can be found in~\cite{frieman1995prl,carroll1998prl, caldwell1998prl, sahni2000ijmpd, urena2000prd, carroll2001lrr, peebles2003rmp, copeland2005prd, martin2008mpla, durrive2018prd}. Other well-known options include Holographic Dark Energy~\cite{li2004plb,pavon2006aip,zimdahl2007cqg}, Chaplygin gas~\cite{kamenshchik2001plb, bilic2002plb, bento2002prd}, phantom field~\cite{chiba2000prd2,caldwell2002plb, carroll2003prd}, quintom model~\cite{feng2005plb,cai2007plb1} (where $w$ evolve to mimic the phantom fluid) among many others. There are excellent reviews~\cite{sahni2002cqg, tsujikawa2013cqg, sami2016ijmpd, brax2018rpp} that summarise the merits and problems of these candidates.

Over the last decade, the availability of high precision data from various surveys has suggested that $w=-1.03\pm 0.03$ within the $95\%$ confidence level~\cite{planck2018cp}, consistent with a cosmological constant. One is tempted to conclude that the cosmological constant as dark energy with cold dark matter (\lcdm) is by far the most suitable model that describes the evolution of the Universe at the present epoch. But the \lcdm model is plagued with problems like the fine-tuning problem~\cite{sahni2002cqg} and the coincidence problem~\cite{steinhardt2003jstor,velten2014epjc}. The fine-tuning problem is that the initial conditions are needed to be set to an exact value so that the cosmological constant term dominates at the current epoch. The coincidence problem is related to the question why the energy densities of dark matter and dark energy are of the same order of magnitude at the present epoch. These problems in the \lcdm model has forced us to look for other candidates that can drive the acceleration. A scalar field rolling down a slowly varying potential not only gives rise to acceleration but also alleviates the cosmological coincidence problem. Such a scalar field, dubbed as `quintessence', has been studied extensively in the literature~\cite{peebles1988apjl, ratra1988prd, ferreira1997prl, copeland1998prd,ferreira1998prd, wetterich1988npb, efstathiou1999mnras, kim1999jhep, zlatev1999prl, steinhardt1999prd, brax1999plb, brax2000prd1,barreiro2000prd, sahni2000prd, albrecht2000prl, dodelson2000prl, wang2000aj, sen2002plb, amendola2006prd, dimopoulos2017jcap, mishra2017jcap, nandan2018prd}. 

The `tracking' model was first introduced by Ratra \& Peebles~\cite{ratra1988prd} and Peebles \& Ratra~\cite{peebles1988apjl}. The idea was to resolve the fine-tuning problem, long before the discovery of the present accelerated expansion of the Universe. Some `scaling' models, which alleviates the fine-tuning problem, were also discussed in~\cite{ferreira1997prl, ferreira1998prd, copeland1998prd, amendola2006prd}, which do not, however, drive the present acceleration. Zlatev \etal~\cite{zlatev1999prl} and Steinhardt \etal~\cite{steinhardt1999prd} later utilised a modified version to incorporate the accelerated expansion. These `tracking' models can resolve the fine-tuning problem and coincidence problem but cannot give rise to the acceleration with $w=-1$. The values of $w$ that can be obtained are $w=-0.6$~\cite{wang2000aj}, $w=-0.8$~\cite{efstathiou1999mnras}, $w\approx-0.82$~\cite{brax2000prd1}, $w<-0.8$~\cite{zlatev1999prl,barreiro2000prd} to mention a few. Thus, none of the \lcdm model and the quintessence models appear to be a complete solution, they are rather complementary to each other. One should therefore look for a model that will have the virtues of both the \lcdm and a quintessence but will be devoid of the flaws. A `tracking' quintessence model with an inverse power law potential, however, was shown to be consistent with observational data sets~\cite{zhai2017aj,park2018aj,ooba2019ass,park2020prd}. The inverse power law potential with a dynamical dark energy gives $w = - 1.03 \pm 0.07$ within the $68.27\%$ confidence limit~\cite{ooba2019ass}, in agreement with the recent observations. However, the present model is different from that discussed in~\cite{efstathiou1999mnras, zlatev1999prl, barreiro2000prd, wang2000aj, brax2000prd1, zhai2017aj, park2018aj, bag2018jcap, ooba2019ass, park2020prd} and yields $w=-1$ at the present epoch independent of the model parameters.

To construct a model without the problem of fixing the initial conditions, the `scaling' potential or the `tracking' potential is a natural choice. However, the scaling solution does not drive an acceleration, whereas the best-known tracking potentials cannot give the observationally preferred value of $w \simeq -1$. In the present work, we introduce a scalar field model with a potential such that it will have an accelerated expansion with an equation of state at the present epoch similar to that given by \lcdm and the current dark energy density comparable to that of dark matter independent of the initial conditions. We engineer the model such that the scalar field $\vphi$ is subdominant as a tracking dark energy at early times and start dominating as a cosmological constant in the recent past driving the acceleration. The presence of a scalar field from early times will have its imprints on the growth of perturbations and hence on the large scale structures of the Universe. The scalar field will evolve through the history of the Universe, and unlike \lcdm, will have fluctuations similar to the other matter components. These fluctuations will affect the formation of structures~\cite{abramo2009prd} and can also cluster on their own~\cite{mehrabi2015mnras, batista2013jcap}. Thus, structure formation will help break the degeneracy between the \lcdm model and our scalar field model ($\vphi$CDM). This work aims to investigate the perturbation in such a dark energy model and look for the distinguishing features from the standard \lcdm model. The present work is not an attempt to constrain the model parameters with the observational datasets but rather to bring out the characteristic features of the model by solving the perturbation equations. It must be mentioned that the motivation of this work is not to unify inflation and dark energy and we will consider the evolution of the $\vphi$CDM long after the completion of inflation. 

The paper is organised as follows. We start with a brief discussion on the scalar field model and introduce the potential in Sect.\ \ref{sec:scf}. In Sect.\ \ref{sec:per}, starting from the relevant perturbation equations we discuss the evolution of the density contrasts along with the CMB temperature fluctuation, matter power spectrum, linear growth rate and $\fsg$. Finally, in Sect.\ \ref{sec:sum}, we summarise and discuss the conclusive results that we arrive at.

\section{The scalar field model} \label{sec:scf}

We consider a homogeneous and isotropic Universe with spatially flat constant time hypersurface, described by the well-known Friedmann-Lema\^itre-Robertson-Walker (FLRW) metric as,
\begin{equation}\label{metric}
ds^2= a^2(\tau)\paren*{- d \tau ^2+\delta_{ij} d x^i dx^j},
\end{equation}
where $a(\tau)$ is the scale factor and the conformal time $\tau$ is related to the cosmic time $t$ as $a^2 d\tau^2 = dt^2$. The Universe is filled with non-interacting fluids, namely photons ($\gamma$), neutrinos ($\nu$), baryons ($b$), cold dark matter ($c$) and a scalar field ($\vphi$) with a potential $\Vp$ acting as dark energy. The Friedmann equations are given as
\begin{eqnarray}
3 \cH^2 &=& -a^2 \kappa \sum_{i} \rho_{i}~,\\ \label{fd1}
\cH^2+ 2 \cH^\prime &=& a^2 \kappa \sum_{i} p_{i}~, \label{fd2}
\end{eqnarray}
where $\kappa=8 \pi G_N$ ($G_N$ being the Newtonian Gravitational constant), $\cH(\tau)= \frac{a^\prime}{a}$ is the conformal Hubble parameter and prime $(^\prime)$ denotes the derivative with respect to the conformal time. 
The energy density and pressure of each component are respectively $\rho_{i}$ and $p_{i}$, where $i = \gamma, \nu, b, c, \vphi$. The equation of state (EoS) parameter is given as $w_{i} = \frac{p_{i}}{\rho_{i}}$. For the photons and neutrinos, $w_{\gamma}=w_{\nu}=1/3$\,, for baryons and CDM, $w_{b} = w_{c} = 0$\,. For the scalar field, $\rphi = \frac{1}{2 a^{2}}\vphi^{\prime \,2} +\Vp$ and $\pphi = \frac{1}{2 a^{2}}\vphi^{\prime \,2} -\Vp$ and the EoS parameter is given by 
\begin{eqnarray}
\wphi = \frac{\pphi}{\rphi}~=~ \frac{\frac{1}{2 a^{2}}\vphi^{\prime \,2} -\Vp}{\frac{1}{2 a^{2}}\vphi^{\prime \,2} +\Vp}~=~ 1-\frac{2 \,\Vp}{\rphi}~. \label{eq:wphi}
\end{eqnarray}
The Klein-Gordon equation can be obtained as a consequence of the Bianchi identities as
\begin{equation} \label{eq:kg1}
\vphi^{\prime \prime} + 2 \cH \vphi^{\prime} + a^{2} \frac{d V}{d \vphi}= 0~.
\end{equation}

It is clear from the expression (\ref{eq:wphi}) that $\wphi$ has an evolution and ranges between $-1 \leq \wphi \leq1$ for a real scalar field and a positive definite $\Vp$. When the kinetic energy ($E_{K} = \frac{\vphi^{\prime \,2}}{2 a^{2}}$) is dominant with a negligible potential energy ($E_{P} = \Vp$), the scalar field behaves as a stiff fluid with $\wphi = 1$, and when $E_{P}$ dominates with a negligible $E_{K}$, it gives rise to a cosmological constant with $\wphi = -1$. Thus, the behaviour of the scalar field and hence the evolution of the Universe depends on the form of the potential. For the recent accelerated expansion of the Universe, the scalar field at late time should roll sufficiently slowly along the potential such that $E_{K} \ll E_{P}$. 

\begin{figure}[!h]
        \centering
\floatbox[{\capbeside\thisfloatsetup{capbesideposition={right,center},capbesidewidth=.5\linewidth}}]{figure}[\FBwidth]
{\caption{Plot of the potential $\Vp$ in units of $\mbox{Gev}^{4}$ against $\vphi/\kappa$ with $V_{0} = 2.510 \times 10^{-47}~\mbox{Gev}^{4}$ and $\lambda = 14.8$ (dashed line), $\lambda = 15.2$ (solid line) and $\lambda = 15.6$ (dashed-dot line). Changing $V_{0}$ will change $\Omega_{\vphi\,0}$.}\label{im:pot}}
{\includegraphics[width=\linewidth]{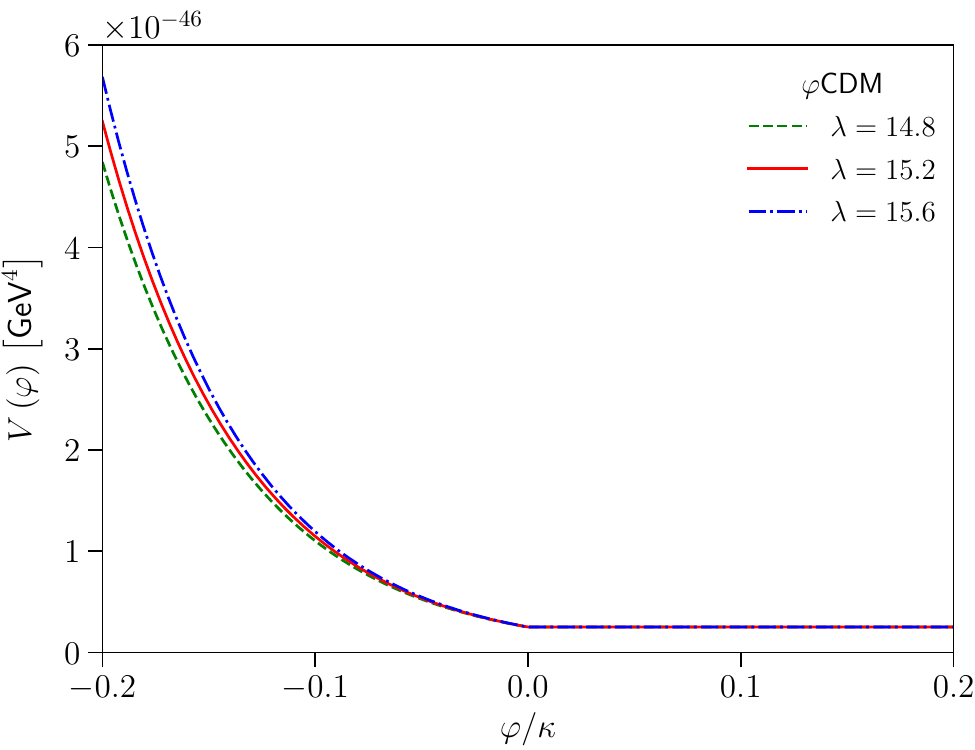}}
\end{figure}

We construct the potential such that the scalar field behaves as a quintessence field in the past and a cosmological constant in the present. We consider the potential as the sum of an exponential potential and a constant potential, shown in Fig.\ (\ref{im:pot}). The potential is written as,
\begin{equation}
\Vp = V_{0}\, e^{-\lambda \kappa \vphi} \Theta\paren*{-\vphi} + V_{0}\, \Theta\paren*{\vphi}~, \label{eq:pot}
\end{equation}
where $V_{0}$ is a constant and $\Theta\paren*{\vphi}$ is the Heaviside theta defined as
\begin{equation}
\Theta\paren*{\vphi} = \begin{cases}
0 & \vphi < 0, \\
1 & \vphi \geq 0.
\end{cases}
\end{equation}
The potential given in Eqn.\ (\ref{eq:pot}) is continuous. In the exponential part, the scalar field tracks the evolution of the dominant background fluid with $\wphi = w_{D}$ and $\Ophi = 3\paren*{1+w_{D}}/\lambda^{2}$ with the condition $\lambda^{2} > 3\paren*{1+w_{D}}$, $w_{D}$ being the EoS parameter of the background fluid and $\Ophi$ is the energy density parameter defined as $\frac{\rho_{\vphi}}{3\,H^2 /\kappa}$. Here, $H$ is the Hubble parameter defined with respect to the cosmic time $t$. This attractor solution is called the `scaling solution', introduced by Ratra \& Peebles in~\cite{ratra1988prd} (see also~\cite{ferreira1997prl, copeland1998prd, ferreira1998prd}). The scalar field then leaves the scaling regime and enters the constant potential regime. The constant part of the potential arrests the fall of the scalar field and it starts to slow-roll and eventually dominate the energy density of the Universe as the cosmological constant. This drives an accelerated expansion at a late time with $\wphi = -1$. 

\begin{figure*}[!h]
\centering
\includegraphics[width=\textwidth]{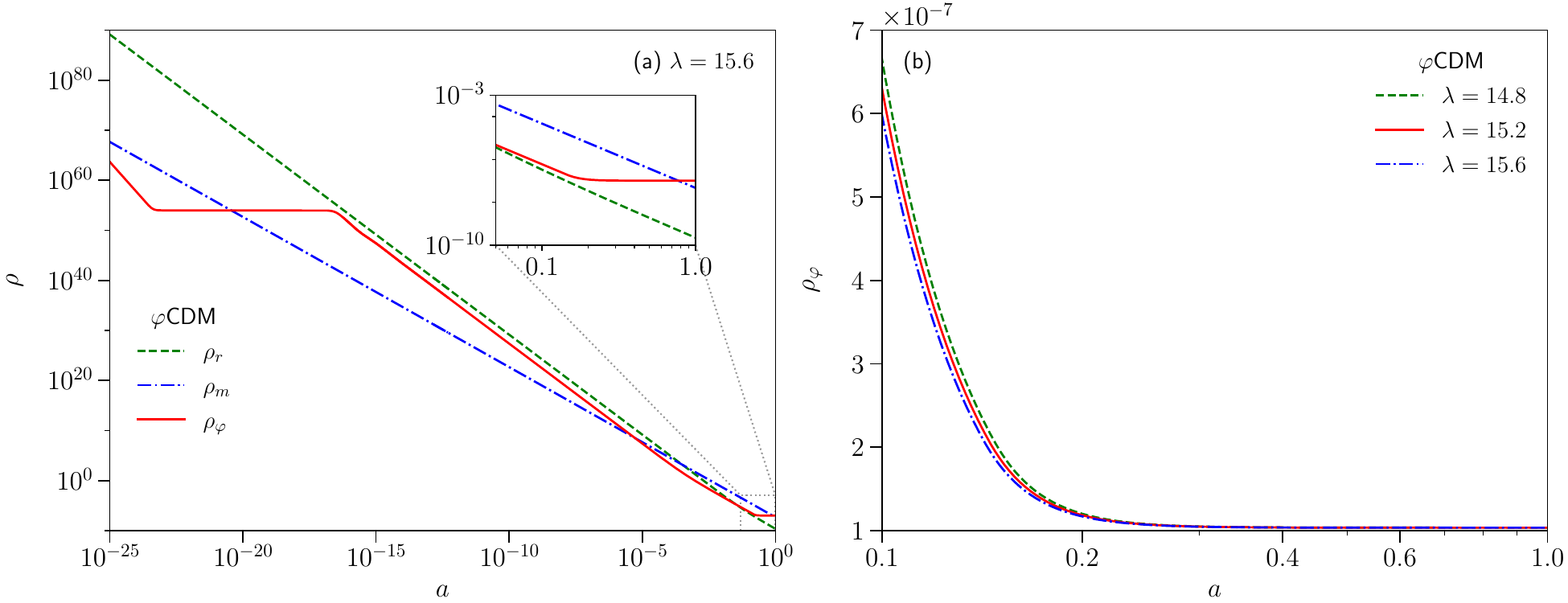}
\caption{(a) Plot of energy density $\rho$ against scale factor $a$ in logarithmic scale where the role of dark energy is played by a scalar field ($\vphi$) in presence of photons ($\gamma$), neutrinos ($\nu$), baryons ($b$), cold dark matter ($c$). For simplicity, only radiation $(r \equiv \gamma+\nu)$ and matter $(m \equiv b+c)$ are shown along with $\vphi$, labelling the model as $\vphi$CDM. The late time evolution of $\rho$ is enlarged in the inset. Only $\lambda =15.6$ is considered here. (b) Plot of $\rphi$ against scale factor $a$ shows that the evolution of $\rho_{\vphi}$ at late time is same for different values of the model parameter $\lambda$ for a fixed value of $V_{0}$.}\label{im:bck1}
\end{figure*}

Fig.\ (\ref{im:bck1}a) shows the variation of the energy density of radiation, $\rho_{r}$ $(r \equiv \gamma+\nu)$, matter, $\rho_{m}$ $(m \equiv b+c)$ and scalar field, $\rphi$, with the scale factor $a$ in logarithmic scale. Before reaching the tracking mode, the scalar field evolves through different regimes as shown in Fig.\ (\ref{im:bck1}a). We integrate the Klein-Gordon equation (\ref{eq:kg1}) numerically, using the potential (\ref{eq:pot}), starting from $a=10^{-25}$ and consider that the initial $\rphi$ is smaller than $\rho_{r}$ and $\rho_{m}$ at that epoch. As $\vphi$ rolls down the potential, $E_{K} \gg E_{P}$ suppressing $\frac{d V}{d \vphi}$ relative to the first two terms in Eqn.\ (\ref{eq:kg1}). $E_{K}$ redshifts as $a^{-6}$ while $E_{P}$ remains constant and $\rphi $ is dominated by $E_{K}$. This kinetic-dominated regime is followed by the potential-dominated regime, where $\vphi$ rolls down very slowly making $\vphi^{\prime \prime}$ inconsequential. In this regime $\rphi$ is determined by $\Vp$ and becomes flat as $\vphi$ hardly evolves. When the dominant background, $\rho_{r}$ in this case, reaches the flat attractor value, they start evolving together. Thereafter, $\rphi$ tracks $\rho_{r}$ and subsequently $\rho_{m}$, depending on which dominates the background as discussed by Ratra \& Peebles~\cite{ratra1988prd}. For a single component background along with the scalar field, such as pure radiation and pure matter, similar results can be obtained analytically as well~\cite{ratra1988prd,steinhardt1999prd,brax2000prd2}. Later, when the constant potential $V_{0}$ takes over, the scalar field energy density, $\rphi$ behaves like the cosmological constant. It should be noted that this transition is instantaneous as it is implemented by a step function. Figure (\ref{im:bck1}b) confirms that the cosmological constant like behaviour is ensured for any value of the parameter $\lambda$ for a given value of $V_{0}$.
 
\begin{figure*}[!h]
        \centering
            \subfloat{\includegraphics[width=.5\linewidth]{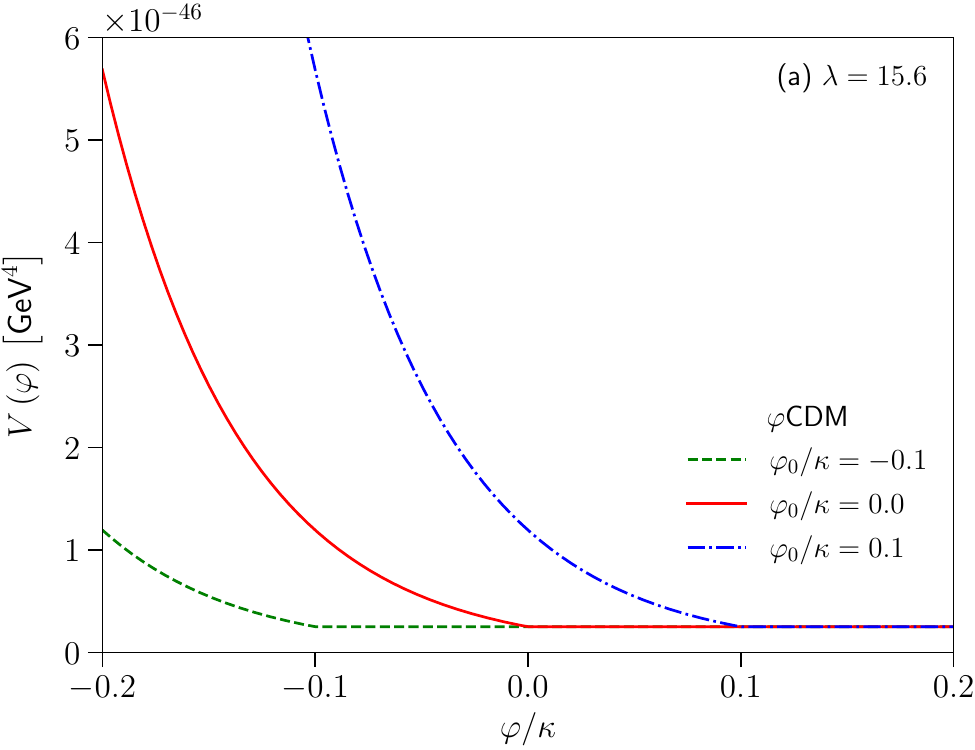}}\hfill
            \subfloat{\includegraphics[width=.5\linewidth]{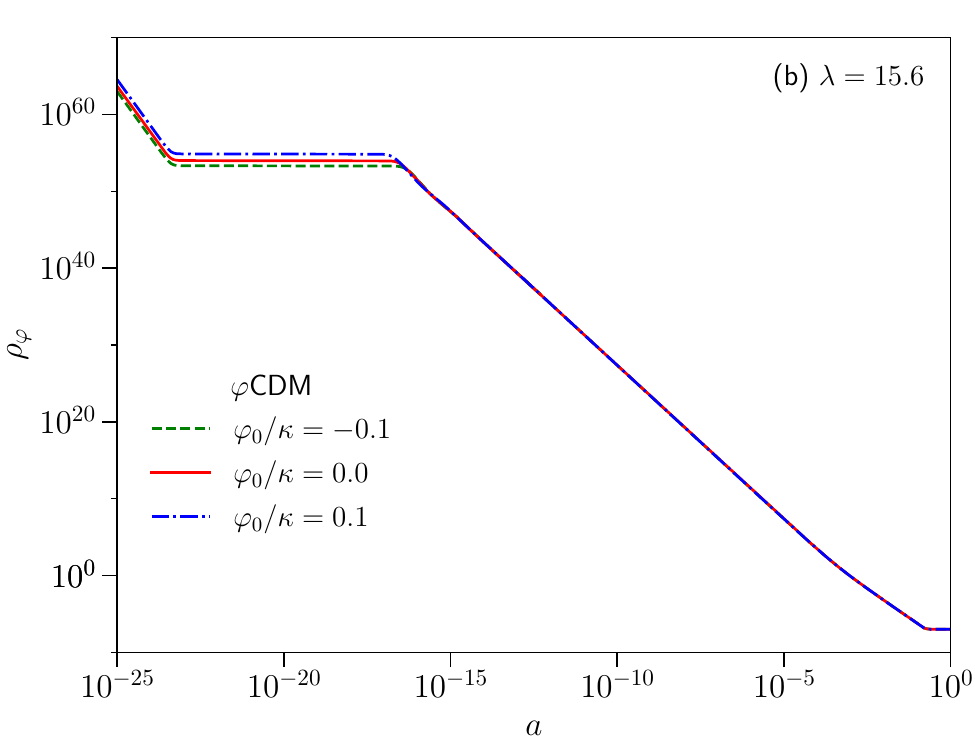}}\hfill
        \caption{(a) Plot of the potential $\Vp$ in units of $\mbox{Gev}^{4}$ against $\vphi/\kappa$ with $V_{0} = 2.510 \times 10^{-47}~\mbox{Gev}^{4}$, $\lambda = 15.6$ and transition at $\vphi_{0}/\kappa = -0.1$ (dashed line), $\vphi_{0}/\kappa = 0.0$ (solid line) and $\vphi_{0}/\kappa = 0.1$ (dashed-dot line). Changing $V_{0}$ will change $\Omega_{\vphi\,0}$. (b) Plot of $\rphi$ against scale factor $a$ shows that the evolution of $\rho_{\vphi}$ is different only at early times for different values of transition point $\vphi_{0}$ for fixed values of $V_{0}$ and $\lambda$.}\label{im:bck4}
\end{figure*}

The advantages of this potential (\ref{eq:pot}) are that at late time $\wphi =-1$, irrespective of the model parameters, $\lambda$ and $V_{0}$, or initial conditions and that the fraction of dark energy density present today, $\Omega_{\vphi0}$ depends on the height of the slow-roll region, $V_{0}$. It deserves mention that $V_{0}$ is not a free parameter but is fixed by the other model parameters like the $\Omega_b h^2$, $\Omega_c h^2$, $H_{0}$, such that $\Omega_{\vphi0}$ matches the observed value of $\Omega_{\Lambda}$ ($\sim 0.6847$)~\cite{planck2018cp}. The $\Theta$ function switches off the effect of the exponential potential in the constant potential part so that the scalar field is dominated completely by the $E_{P}$ after leaving the scaling region. It deserves mention that the value of $\vphi$ for the transition from exponential to constant potential is a free parameter. The effect in the evolution of the scalar field due to the change in the transition value can be seen from Fig.\ (\ref{im:bck4}), where we consider three examples. When the transition point is shifted from zero to $\vphi_{0}$, the exponential part of the potential, Eqn.\ (\ref{eq:pot}) changes as $V_{0}\,e^{-\lambda \kappa \paren*{\vphi-\vphi_{0}}}$ to accommodate for the continuity of $\Vp$. The change in the steepness of the exponential potential changes the evolution of the scalar field before it behaves as a tracking field. Once it starts to track the dominant background components, its evolution remains unaffected by the change in the transition point, $\vphi_{0}$. So without loss of much of generality, we define our potential with $\vphi_{0}=0$.

The constraint on the parameter $\lambda$ comes from Big Bang Nucleosynthesis (BBN) condition~\cite{wetterich1988npb,copeland1998prd,ferreira1998prd},
\begin{equation}
\Ophi\paren*{a\sim 10^{-10}} \lesssim 0.09.
\end{equation}
It should be noted that in all the subsequent discussion, the scale factor, $a$, is scaled so that its present value, $a_0=1$. Considering $V_{0} = 2.510 \times 10^{-47}~\mbox{Gev}^{4}$ and $\lambda = 15.6$ with the parameter values listed in Table \ref{tab:bck} gives $\Ophi\paren*{a\sim 10^{-10}} = 0.01642$ and $\Ophi\paren*{a = 1} = 0.6840$. The parameter values listed in Table \ref{tab:bck}, are taken from the latest data release in 2018 of the \Planck collaboration~\cite{planck2018cp} (\Planck 2018, henceforth) and are based on the fiducial spatially flat \lcdm model. For our calculation we have considered $\vphi_{i} = -\frac{8.99}{\kappa}$ at $a_{i} = 10^{-25}$. It turns out that $\vphi = 0$ at $a = 0.14237$ (for the values of $V_{0}$ and $\lambda$ chosen), where the potential changes its role from a scaling potential to effectively a cosmological constant. The dimensionless density parameter, $\Omega_i$ is given by $\frac{\rho_i}{3\,H^2 /\kappa}$ where the suffix $i$ stands for the $i$-th component. The dimensionless Hubble parameter at the present epoch is defined as $h = \frac{H_0}{100 \hskip1ex \footnotesize{\mbox{km s}^{-1} \mbox{Mpc}^{-1}}}$. Figure (\ref{im:bck1}) is obtained by solving the Klein-Gordon Eqn.\ (\ref{eq:kg1}) numerically with the potential (\ref{eq:pot}) using these parameter values. For the study of detailed dynamics of the scalar field during tracking region we refer to~\cite{ratra1988prd, ferreira1997prl, copeland1998prd, ferreira1998prd,  steinhardt1999prd, brax2000prd2}.
\begin{table}[!h]
\begin{center}
\caption{\label{tab:bck}
Values of background parameters from the \Planck 2018 collaboration.
}
\begin{tabular}{cc}
\hline \hline
Parameter& \hspace{24ex} Value\\
\hline
\rule[-1ex]{0pt}{2.5ex}$\Omega_b h^2$ & \hspace{24ex} $0.0223828$ \\
\rule[-1ex]{0pt}{2.5ex}$\Omega_c h^2$ & \hspace{24ex} $0.1201075$ \\
\rule[-1ex]{0pt}{2.5ex}$H_0 \left[ \mbox{km s}^{-1} \mpci \right]$ & \hspace{24ex} $67.32117$ \\
\hline
\hline
\end{tabular}
\end{center}
\end{table}

\begin{figure*}[!h]
\centering
\includegraphics[width=\textwidth]{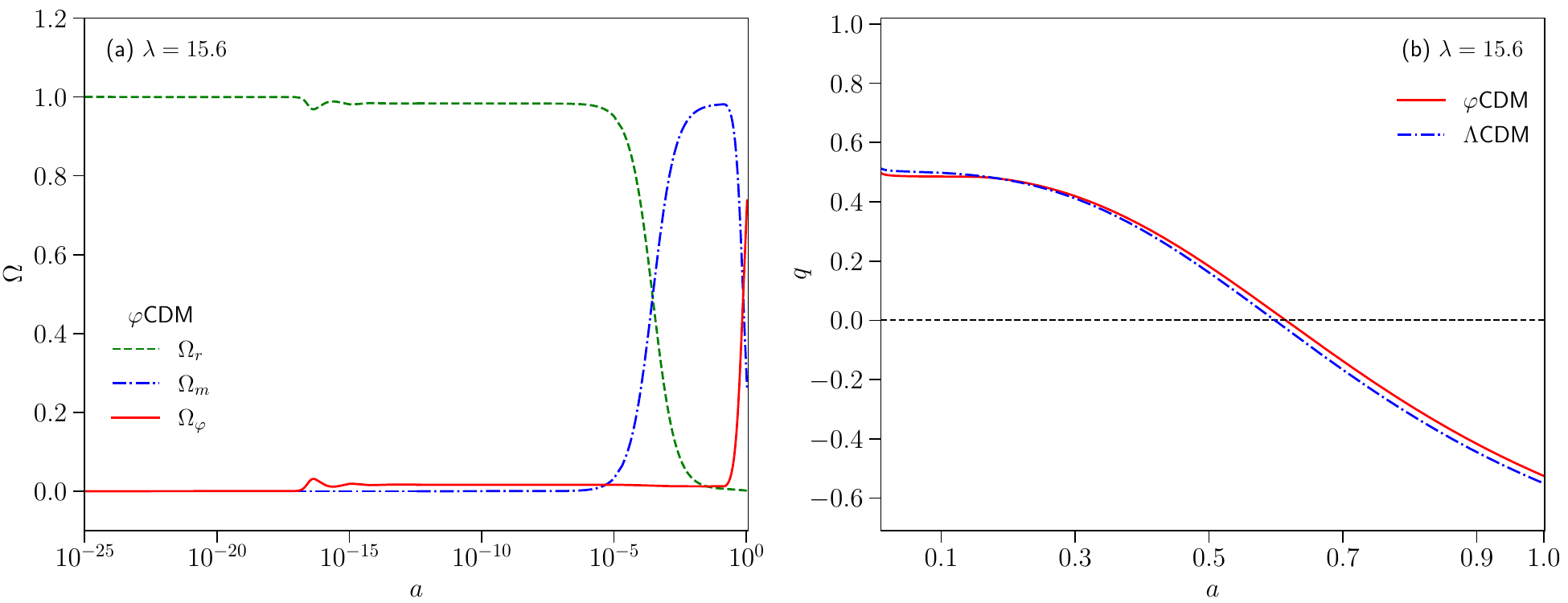}
\caption{(a) Plot of density parameter $\Omega$ against scale factor $a$ in logarithmic scale. For simplicity, only radiation $(r \equiv \gamma+\nu)$ and matter $(m \equiv b+c)$ are shown along with $\vphi$. (b) Plot of deceleration parameter $q$ against scale factor $a$ for $\vphi$CDM (solid line) and \lcdm (dashed-dot line). Only $\lambda =15.6$ is considered here.}\label{im:om}
\end{figure*}

The evolution of the energy density parameters, $\Omega \paren*{\equiv \Omega_{i}}$ of radiation $(r \equiv \gamma+\nu)$, matter $(m \equiv b+c)$ and scalar field $(\vphi)$ with the scale factor $a$, in logarithmic scale, are shown in Fig.\ (\ref{im:om}a) and that of the deceleration parameter $q = -\paren*{\frac{a\, a^{\prime \prime}}{a^{\prime \,2}}-1}$ with $a$ in Fig.\ (\ref{im:om}b) for $\lambda=15.6$. Figure (\ref{im:om}) shows that the evolution dynamics of the Universe is different from the \lcdm model even though $\wphi=-1$ at the present epoch. The two models are qualitatively very similar, but not really overlapping. For the scalar field model, henceforth called $\vphi$CDM, the accelerated expansion starts at a little higher value of $a$ compared to the \lcdm model.

\section{The perturbations}\label{sec:per}

The scalar field model given by Eqn.\ (\ref{eq:pot}) can have fluctuations and thereby affect the evolution of perturbations of other components. The scalar perturbation equations in synchronous gauge are considered in the present work and the differential equations are solved using the suitably modified version of the publicly available Boltzmann code \camb\footnote{\href{https://camb.info}{https://camb.info}}~\cite{lewis1999bs}. To study the dependence of the fluctuations on the model parameters, we used different values of $\lambda$ keeping $V_{0}$ constant (varying $V_{0}$ will change $\Omega_{\vphi\,0}$).

\subsection{Effect on density perturbation}\label{sec:dp}

The scalar perturbation of the FLRW metric takes the form~\cite{valiviita2008jcap} 
\begin{equation} \label{eq:metric2}
\begin{split}
ds^2=a^2\paren*{\tau} & \left\{ -\paren*{1+2\phi}d\tau^2+2\,\partial_iB \,d\tau \,dx^i +\left[\paren*{1-2\psi}\delta_{ij}+2\partial_i\partial_jE\right]dx^idx^j \right\},
\end{split}
\end{equation}
where $\phi, \psi, B, E$ are gauge-dependent functions of both space and time. In synchronous gauge, $\phi=B=0$, $\psi=\eta$ and $k^2 E=-\msh/2-3\eta$, where $\eta$ and $\msh$ are the synchronous gauge fields defined in the Fourier space and $k$ is the wavenumber~\cite{ma1995aj}. The perturbation equations in the matter sector in the Fourier space are
\begin{eqnarray}
\delta_{i}^\prime+ k \,v_{i} +\frac{\msh^\prime}{2} &=& 0,\label{e2dm} \\ 
v_{i}^\prime+\cH v_{i}&=& 0, \label{m2dm}
\end{eqnarray}
where $\delta_{i} = \delta \rho_{i}/\rho_{i}$ is the density contrast and $v_{i}$ is the peculiar velocity of $i$-th $\paren*{i= b,c}$ fluid. Assuming there is no momentum transfer in CDM frame, $v_{c}$ is set to zero. For the details of this set of equations, we refer to the works~\cite{kodama1984ptps,mukhanov1992pr, ma1995aj, malik2003prd}. The perturbation $\delta \vphi$ in the scalar field has the equation of motion~\cite{cembranos2016jhep}
\begin{equation} \label{eq:kg2}
\delta \vphi^{\prime \prime} + 2 \cH \delta \vphi^{\prime} + k^{2}\delta \vphi +a^{2}~\frac{d^{2} V}{d \vphi^{2}} \delta \vphi + \frac{1}{2} \vphi^{\prime} \msh^{\prime}= 0,
\end{equation}
in the Fourier space with wavenumber $k$. The perturbation in energy density $\delta \rphi$ and pressure $\delta \pphi$ are given as
\begin{eqnarray}
\delta \rphi &=& -\delta T_{0 \paren*{\vphi}}^{0}~=~ \frac{\vphi^{\prime} \delta \vphi^{\prime}}{a^{2}}+\delta \vphi \frac{d V}{d \vphi} , \label{eq:pe2}\\
\delta T_{0\paren*{\vphi}}^{j} &=& - \frac{ \im k_{j}\, \vphi^{\prime}\, \delta \vphi}{a^{2}}, \hspace{0.6cm} \im \equiv \sqrt{-1} \label{eq:pv2}\\
\delta \pphi \delta^{i}_{j}&=& \delta T_{j\paren*{\vphi}}^{i}~=~ \paren*{\frac{\vphi^{\prime} \delta \vphi^{\prime}}{a^{2}}-\delta \vphi \frac{d V}{d \vphi}}\delta^{i}_{j}\label{eq:pp2},
\end{eqnarray}
when expanded in the Fourier space. Here $\delta T^{\mu}_{\nu\paren*{\vphi}}$ is the perturbed stress-energy tensor of the scalar field.

For an adiabatically expanding Universe, the square of sound speed is $\csf = \pphi^{\prime}/\rphi^{\prime}$. Using the Klein-Gordon Eqn.\ (\ref{eq:kg1}), the square of adiabatic sound speed~\cite{martin1998prd,brax2000prd2} for the scalar field reads as
\begin{equation}
\csf = -\frac{1}{3}-\frac{2 \vphi^{\prime \prime}}{3 \cH \vphi^{\prime}} = 1+\frac{2 a^{2}}{3 \cH \vphi^{\prime}} \frac{d V}{d \vphi}.\label{eq:sound}
\end{equation}
In order to solve the perturbation Eqn.\ (\ref{eq:kg2}), the second derivative of the potential is written in terms of the square of sound speed, $\csf$ as
\begin{equation}
\frac{d^{2} V}{d \vphi^{2}} = \frac{3}{2} \frac{\cH^{2}}{a^{2}}\left[\frac{c_{s,\vphi}^{2 \,\,\prime}}{\cH}-\frac{1}{2}\paren*{\csf-1}\paren*{3 \csf+5}+\frac{\cH^{\prime}}{\cH}\paren*{\csf-1}\right].\label{eq:v2}
\end{equation}
The square of sound speed, $\csf$ is constant in the different phases of evolution, e.g.\ in the scaling regime $\csf = \wphi = w_{D}$ and in the slow-roll regime $\csf = 1$. We shall henceforth take it to be described by Eqn.\ (\ref{eq:sound}) but neglect its derivative, $c_{s,\vphi}^{2 \,\,\prime}$~\cite{brax2000prd2} in Eqn.\ (\ref{eq:v2}). The perturbation Eqns.\ (\ref{e2dm}) and (\ref{m2dm}) are solved along with Eqns.\ (\ref{eq:kg2}), (\ref{eq:pe2}) and (\ref{eq:pv2}) with adiabatic initial conditions and $k = \left[1.0, 0.1, 0.01 \right] h$ $\mpci$ using \camb.

\begin{figure*} [!h]
 \centering
\includegraphics[width=0.95\textwidth]{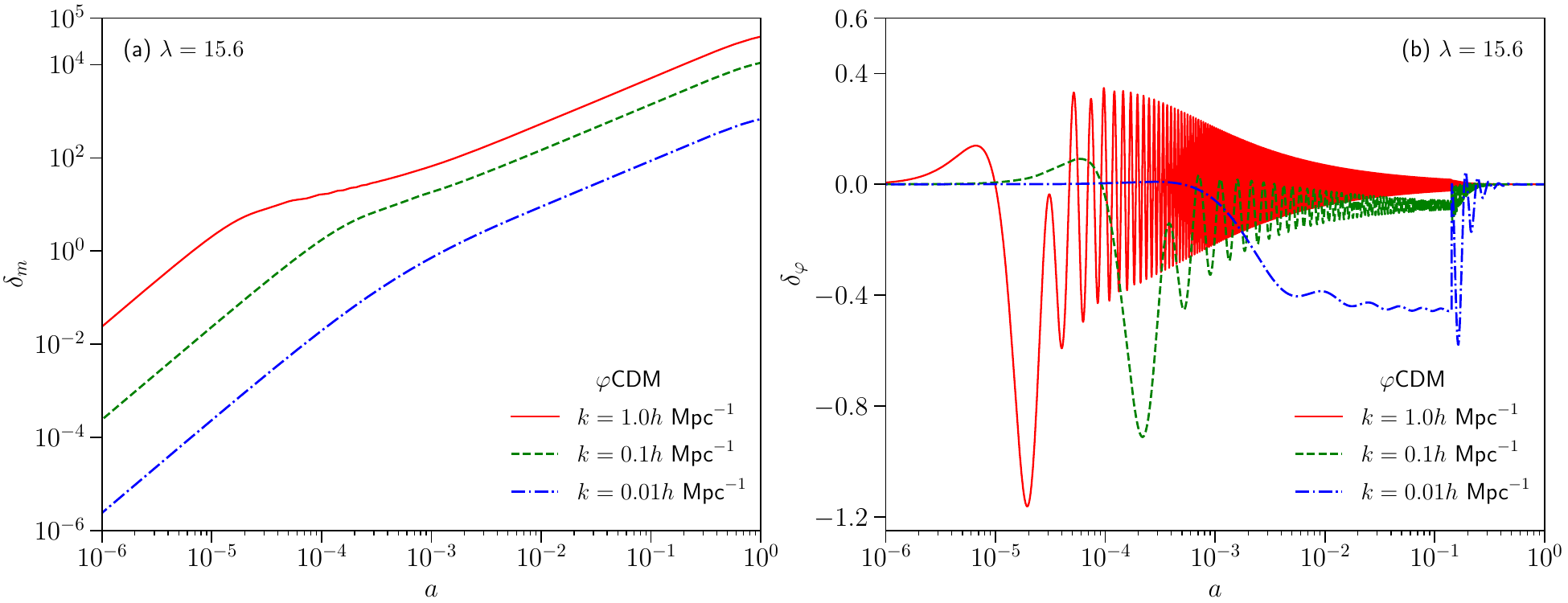}
\caption{(a) Plot of the matter density contrast $\delta_{m}$ against $a$. Both the axes are in logarithmic scale. (b) Plot of scalar field density contrast $\delta_{\vphi}$ against $a$. In (b), only $a$ in logarithmic scale. The solid line represents $k = 1.0\,h$ $\mpci$, dashed line represents $k = 0.1\,h$ $\mpci$ and dashed-dot line represents $k = 0.01\,h$ $\mpci$ with $\lambda = 15.6$.}\label{im:delta2}
\end{figure*}

\begin{figure}[!h]
        \centering
\floatbox[{\capbeside\thisfloatsetup{capbesideposition={right,center},capbesidewidth=.5\linewidth}}]{figure}[\FBwidth]
{\caption{\textbf{Upper Panel} : Plot of the matter density contrast $\frac{\delta_{m}}{\delta_{m\,0,\,\scriptsize{\Lambda\text{CDM}}}}$ against $a$ in logarithmic scale for $\vphi$CDM with $\lambda = 15.2$ (solid line with solid circles) and $\lambda = 15.6$ (solid line) and \lcdm (dashed-dot line) for $k = 0.1\,h$ $\mpci$. The difference in the growth of $\delta_{m}$ for $\vphi$CDM and \lcdm is prominent in the recent past, so the plot starts from $a =10^{-3}$. \textbf{Lower Panel} : Plot of the fractional growth rate relative to the \lcdm model. The fractional growth rate is defined as $\frac{\Delta \delta_{m}}{\delta_{m,\,\scriptsize{\Lambda\text{CDM}}}} = \paren*{1- \frac{\delta_{m,\,\scriptsize{\vphi\text{CDM}}}}{\delta_{m,\,\scriptsize{\Lambda\text{CDM}}}}}$.}\label{im:delta6}}
{\includegraphics[width=\linewidth]{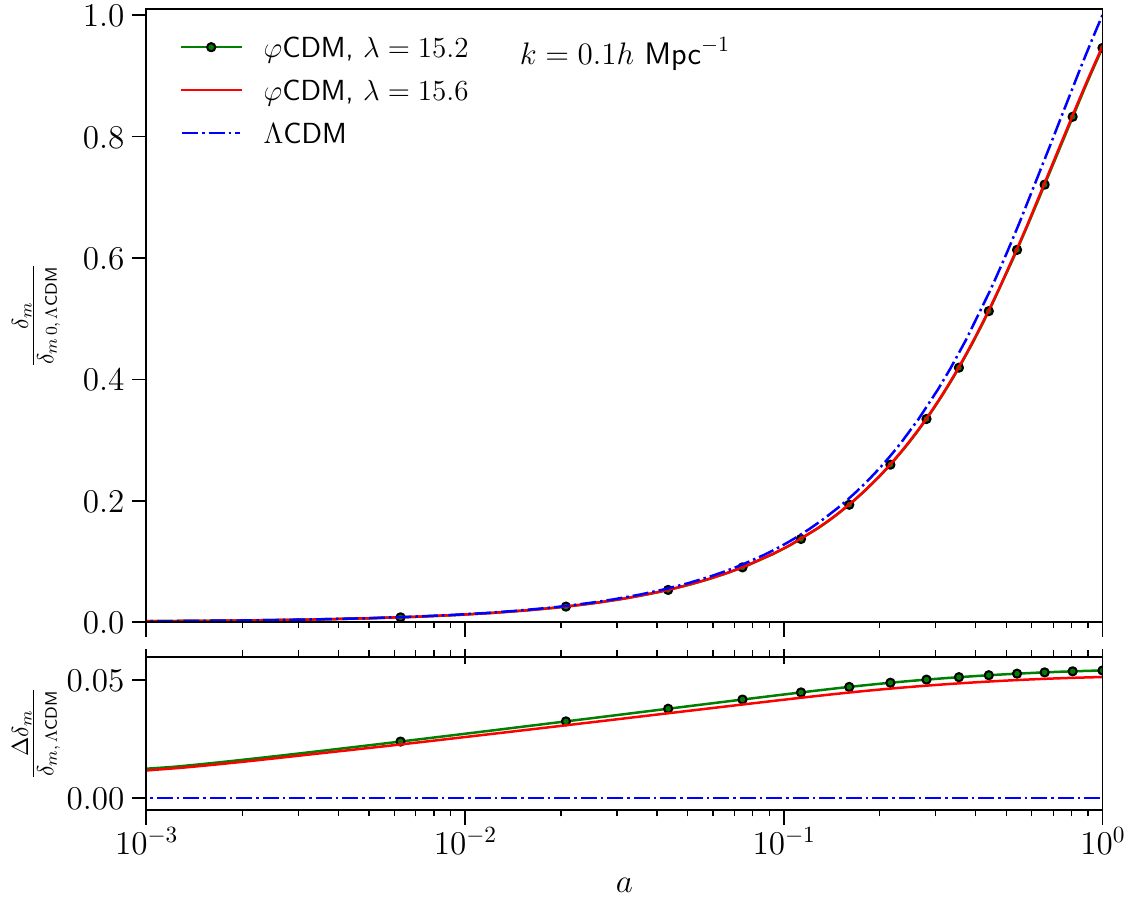}}
\end{figure}

Figure (\ref{im:delta2}a) shows the variation of the density contrast, $\delta_{m} = \delta \rho_{m}/\rho_{m}$ for the cold dark matter ($c$) together with the baryonic matter ($b$) and Fig.\ (\ref{im:delta2}b) shows the variation of the density contrast $\delta_{\vphi} = \delta\rho_{\vphi}/\rho_{\vphi}$ of the scalar field against $a$ in logarithmic scale for $k = \left[1.0, 0.1, 0.01 \right] h$ $\mpci$. In the matter dominated era, the modes of $\delta_{m}$ grow in a very similar fashion. The modes of $\delta_{\vphi}$ oscillate rapidly with decreasing amplitude after entering the horizon. Figure (\ref{im:delta6}) shows the evolution of the matter density contrast $\delta_{m}$, for $\vphi$CDM and \lcdm. For a better comparison, $\delta_{m}$ for both the models have been scaled by $\delta_{m0} = \delta_{m}\paren*{a=1}$ of \lcdm. It can be seen that there is a difference in the growth of $\delta_{m}$ in the two models ($\vphi$CDM and \lcdm). To distinguish the effect of the parameter, $\lambda$, of the present potential with the \lcdm model, we have shown the fractional matter density contrast, $\frac{\Delta \delta_{m}}{\delta_{m,\,\scriptsize{\Lambda\text{CDM}}}} = \paren*{1- \frac{\delta_{m,\,\scriptsize{\vphi\text{CDM}}}}{\delta_{m,\,\scriptsize{\Lambda\text{CDM}}}}}$ in the lower panel of Fig.\ (\ref{im:delta6}). It is clearly seen that, $\delta_{m}$ for $\lambda=15.2$ takes a slightly smaller value compared to that of $\delta_{m}$ for $\lambda=15.6$. Thus, the growth of matter density fluctuation decreases with decrease in the parameter, $\lambda$.


\subsection{Effect on CMB temperature, matter power spectra and $\fsg$}\label{sec:ps}

\begin{figure}[!h]
 \centering
\includegraphics[width=\linewidth]{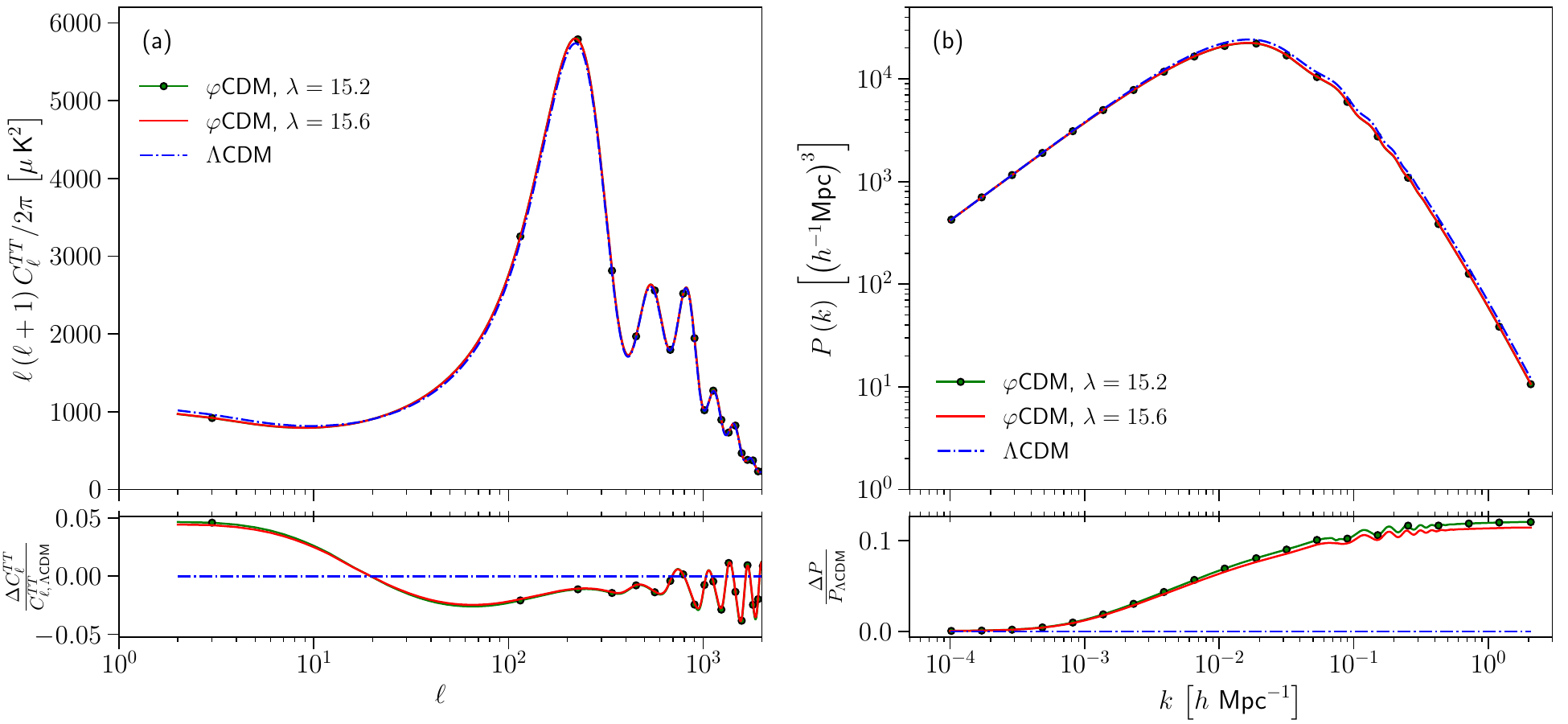}
\caption{\textbf{Upper Panel} : (a) Plot of CMB temperature power spectrum in units of $\mu \mbox{K}^{2}$ with the multipole index $\ell$ in logarithmic scale. (b) Plot of matter power spectrum $P\paren*{k}$ in units of $\paren*{h^{-1}\mpc}^{3}$ with wavenumber $k$ in units of $h\,\mpci$. Both the axes are in logarithmic scales in (b). \textbf{Lower Panel} : Plot of fractional change in the temperature spectrum, $\frac{\Delta C_{\ell}^{TT}}{C_{\ell,\, \scriptsize{\Lambda\text{CDM}}}^{TT}} = \paren*{1- \frac{C_{\ell,\, \scriptsize{\vphi\text{CDM}}}^{TT}}{C_{\ell,\, \scriptsize{\Lambda\text{CDM}}}^{TT}}}$ and the fractional change in matter power spectrum, $\frac{\Delta P}{P_{\scriptsize{\Lambda\text{CDM}}}} = \paren*{1- \frac{P_{\scriptsize{\vphi\text{CDM}}}}{P_{\scriptsize{\Lambda\text{CDM}}}}}$ For both panels, the solid line with solid circles represents $\vphi$CDM with $\lambda = 15.2$ and solid line represents $\vphi$CDM with $\lambda = 15.6$ while the dashed-dot line is for \lcdm at $a=1$.}\label{im:mt}
\end{figure}

For more insight into the effect of the scalar field $\vphi$ on different physical quantities, we look at the CMB temperature spectrum, matter power spectrum and $\fsg$. The CMB temperature power spectrum is given as
\begin{equation}
C_{\ell}^{TT} = \frac{2}{k} \int k^{2} d k \,P_{\zeta}\paren*{k} \Delta^{2}_{T\ell}\paren*{k},
\end{equation}
where $P_{\zeta}\paren*{k}$ is the primordial power spectrum, $\Delta_{T\ell}\paren*{k}$ is the temperature transfer function, $\ell$ is the multipole index and $T$ stands for temperature. For the detail calculation of the CMB spectrum we refer to~\cite{hu1995apj,seljak1996apj}. The matter power spectrum is given as
\begin{equation} \label{power}
P\left(k,a\right)= A_s \,k^{n_s} T^2\left(k\right) D^2\left(a\right),
\end{equation}
where $A_{s}$ is the normalising constant, $n_{s}$ is the spectral index, $T\left(k\right)$ is the matter transfer function and $D\left(a\right)=\frac{\delta_m\left(a\right)}{\delta_m\left(a=1\right)}$ is the normalised density contrast. For the detailed method of calculation we refer to the monograph by Dodelson~\cite{dodelson2003}. The $C_{\ell}^{TT}$ and $P\paren*{k,a}$ are computed numerically using \camb. The values $A_{s} = 2.100549 \times 10^{-9}$ and $n_{s} = 0.9660499$ are taken from the \Planck 2018 data~\cite{planck2018cp}, and hence, depend on the fiducial \lcdm model. Figure (\ref{im:mt}a) shows that the CMB temperature power spectra, $C_{\ell}^{TT}$, are almost independent of the values of the model parameter, $\lambda$. For clarity of the plots only two values of $\lambda$ are given. The presence of the scalar field $\vphi$ decreases the matter content of the Universe slightly during matter domination making the amplitude of first two peaks of the CMB spectra marginally higher than that in the \lcdm model. The scalar field also lowers the low-$\ell$ CMB spectrum through the integrated Sachs-Wolfe (ISW) effect. These features are clear from the lower panel of Fig.\ (\ref{im:mt}a), which shows the fractional change ($=\Delta C_{\ell}^{TT}/C_{\ell,\, \scriptsize{\Lambda\text{CDM}}}^{TT}$) in $C_{\ell}^{TT}$ of the $\vphi$CDM models relative to the \lcdm model; a smaller $\lambda$ produces slightly lower low-$\ell$ modes. A lesser amount of matter leads to a marginally lower matter power spectrum at small scales (Fig.\ (\ref{im:mt}b)), which is clear from the positive fractional change in matter power spectrum, $\Delta P/P_{\scriptsize{\Lambda\text{CDM}}}$, relative to the \lcdm model (lower panel). Both these figures are for the present epoch.
 
To differentiate the $\vphi$CDM and \lcdm decisively, we have studied the linear growth rate, 
\begin{equation} \label{eq:growth_rate}
f\paren*{a}= \frac{d \ln \delta_{m}}{d \ln a}~=~ \frac{a}{\delta_{m}\paren*{a}}\frac{d \delta_{m}}{d \,a}~.
\end{equation}
Observationally the growth rate is measured using the perturbation of the galaxy density $\delta_{g}$, which is related to the matter density perturbations $\delta_{m}$ as $\delta_{g} = b \delta_{m}$, where $b \in \left[1,3\right]$ is the bias parameter. The estimate of the growth rate $f$ is sensitive to the bias parameter, and thus not very reliable. A more dependable observational quantity is the product $f\paren*{a}\se\paren*{a}$~\cite{percival2009mnras}, where $\se\paren*{a}$ 
is the root-mean-square (rms) fluctuations of the linear density field within the sphere of radius $R = 8 h^{-1}\, \mpc$. The rms mass fluctuation can be written as $\se\paren*{a} = \se\paren*{1}\frac{\delta_{m}\paren*{a}}{\delta_{m}\paren*{1}}$, where $\se\paren*{1}$ is the value at $a=1$ (Table \ref{tab:s8}), calculated by integrating the matter power spectrum over all the values of the wavenumber $k$ using \camb. Thus, the combination becomes
\begin{equation} \label{eq:fsigma}
\fsg\paren*{a} \equiv f\paren*{a}\se\paren*{a} = \se\paren*{1}\frac{a}{\delta_{m}\paren*{1}}\frac{d \delta_{m}}{d \,a}~.
\end{equation}

Since $\fsg$ measurements provide a tighter constraint on the cosmological parameters, it will give a better insight into the growth of the density perturbations. We have studied the variation of $f$ and $\fsg$ with redshift $z$ for three different values of $\lambda$. Redshift $z$ is related to the scale factor $a$ as $z = \paren*{\frac{a_0}{a} -1}$, $a_0$ being the present value. The linear growth rate $f$ and $\fsg$ are independent of the wavenumber $k$ for low redshift. As the $\fsg$ analysis is valid for $z \in \left[0,2\right]$, the redshift from $z=0$ to $z=2$ are considered here.
\begin{table}[h!]
\begin{center}
\caption{\label{tab:s8}
Values of $\se$ at $a = 1$ for the $\vphi$CDM and \lcdm models.
}
\begin{tabular}{ccc}
\hline
\hline
Model& \hspace{15ex}$\lambda$&\hspace{15ex} $\se$ \\
\hline
\rule[-1ex]{0pt}{2.5ex}&\hspace{15ex}$14.8$ &\hspace{15ex} $0.7638$ \\
\rule[-1ex]{0pt}{2.5ex}$\vphi$CDM&\hspace{15ex}$15.2$ &\hspace{15ex} $0.7664$ \\
\rule[-1ex]{0pt}{2.5ex}&\hspace{15ex}$15.6$ &\hspace{15ex} $0.7687$ \\
\rule[-1ex]{0pt}{2.5ex}\lcdm&\hspace{15ex} --- &\hspace{15ex} $0.8123$ \\
\hline
\hline
\end{tabular}
\end{center}
\end{table}

\begin{figure*}[!h]
 \centering
\includegraphics[width=\textwidth]{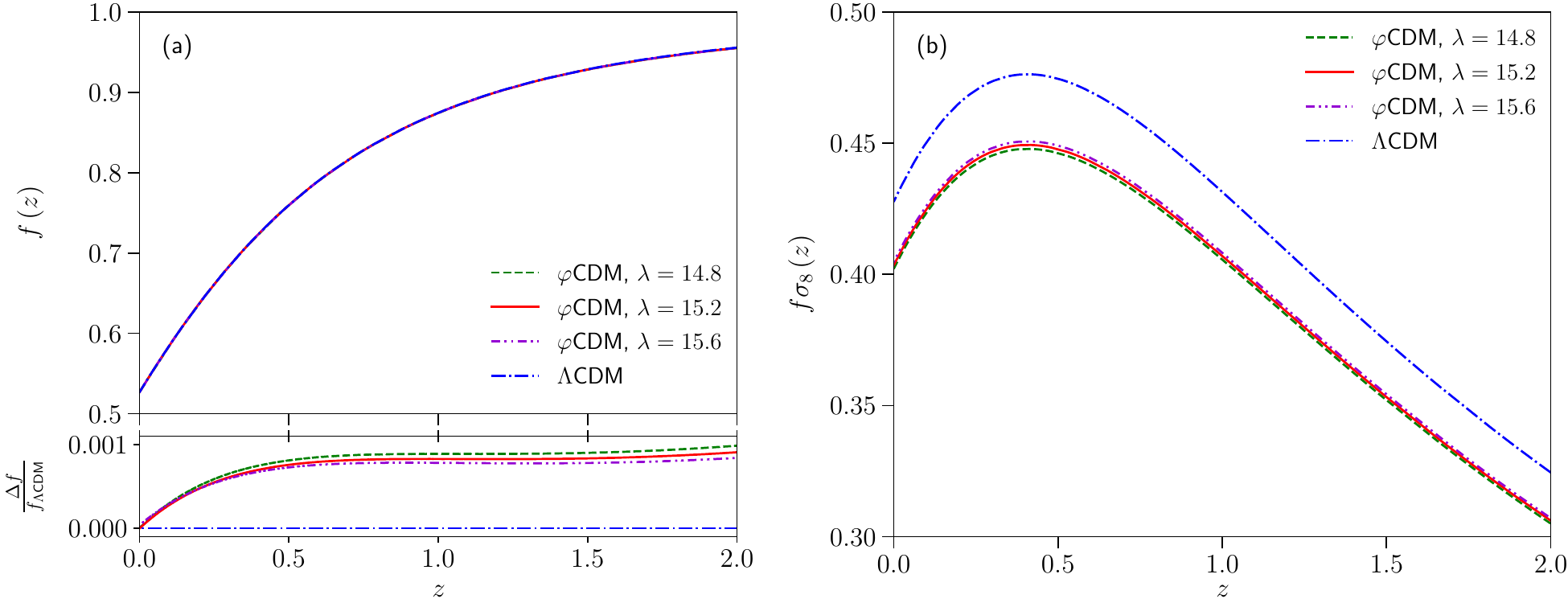}
\caption{(a) Plot of \textbf{Upper Panel} : linear growth rate $f$ and \textbf{Lower Panel} : fractional growth rate, $\frac{\Delta f}{f_{\scriptsize{\Lambda\text{CDM}}}}  = \paren*{1-\frac{f_{\scriptsize{\vphi\text{CDM}}}}{f_{\scriptsize{\Lambda\text{CDM}}}}}$ relative to the \lcdm model. (b) Plot of $\fsg$ against redshift $z$. For all the plots, the dashed line represents $\vphi$CDM with $\lambda = 14.8$, solid line represents $\lambda = 15.2$ and dashed-dot-dot represents $\lambda = 15.6$ while the dashed-dot line is for \lcdm.}\label{im:fs}
\end{figure*}

The linear growth rate $f$ is almost same for all the models at low redshift (Fig.\ (\ref{im:fs}a)). The little change in the growth rate, $f$ due to change in $\lambda$ is visible in the fractional change in growth rate, $\Delta f/f_{\scriptsize{\Lambda\text{CDM}}}$, relative to the \lcdm model (lower panel of Fig.\ (\ref{im:fs}a)). The difference in matter power spectrum is manifested in its amplitude $\se$ as given in Table (\ref{tab:s8}) and hence in $\fsg$ as in Fig.\ (\ref{im:fs}b). It is interesting to note that there is a substantial difference in $\fsg$ for $\vphi$CDM and \lcdm which is not there in the CMB temperature and matter power spectra. Thus, a low $\fsg$ can be said to be the characteristic distinguishing feature of the present $\vphi$CDM model from the \lcdm model. It must be noted that $\lambda$ is chosen in such a way that is compatible with the age of the Universe which is around $13.797 \pm 0.023$ Giga years according to the recent \Planck 2018 data~\cite{planck2018cp}. 

\section{Discussion}\label{sec:sum}

In the present work, we have introduced a scalar field model that will retain the virtues of the \lcdm model without its shortcomings. We have investigated the perturbation in such a dynamical dark energy model that will alleviate the initial condition problem associated with the cosmological constant and attain an EoS parameter $\wphi = -1$ at the present epoch. At early times the scalar field energy density tracks the dominant component of the background fluid and later on starts to roll sufficiently slowly to drive the accelerated expansion of the Universe. A scalar field with an exponential potential at early epoch and a constant potential at late time connected by Heaviside $\Theta$ functions (see Eqn.\ \ref{eq:pot}) appears to serve the purpose. That $\wphi = -1$ for the present epoch is independent of the choice of the model parameters and the present dark energy density parameter, $\Omega_{\vphi\,0}$ is dependent on the height of the constant potential, $V_{0}$.

We have worked out a detail perturbation analysis to differentiate the scalar field model ($\vphi$CDM) with the \lcdm model. We have analysed the physical quantities that can be arrived at by solving the perturbation equations. The linearised scalar perturbations of the FLRW metric in synchronous gauge are studied using our modified \camb. The growth of matter density contrast, $\delta_{m}$ is similar to the \lcdm model and is smaller for the smaller value of $\lambda$. The linear growth rate $f$, which is the logarithmic derivative of $\delta_{m}$ with respect to $a$ is same for both the models. The presence of the scalar field slightly decreases the matter content of the Universe during the evolutionary history. This decrease in matter content is manifested in the matter power spectrum and even more clearly in the evolution of the $\fsg$. Thus, $\fsg$ helps in breaking the degeneracy between the present $\vphi$CDM model and the standard \lcdm. Another interesting result is that the decrease in the rate of clustering decreases the variance of the linear matter perturbation, $\se$. As seen from Table (\ref{tab:s8}), the $\se$ obtained here is more towards the side of the value obtained from the galaxy cluster counts using thermal Sunyaev-Zel'dovich (tSZ) signature~\cite{planck2018cp,zubeldia2019mnras}, $\se = 0.77^{+0.04}_{-0.03} $ rather than the value obtained from \Planck spectrum~\cite{planck2018cp}, $\se = 0.811 \pm 0.006$. It must be mentioned here that as shown in~\cite{banerjee2020xcn,eoin2019plb}, quintessence-CDM model also prefer lower $H_{0}$ compared to the standard \lcdm model. A detailed study of the parameter space is required to confirm if this model can solve the $\se$ tension and prefer a lower value of $H_{0}$. Such an analysis is outside the scope of the present article and will be considered in a separate work. It has already been mentioned that the values of the background parameters ($\Omega_b h^2$, $\Omega_c h^2$, $H_{0}$, $A_{s}$ and $n_{s}$) are fixed to the mean values obtained by the \Planck 2018 collaboration~\cite{planck2018cp}, for the fiducial spatially flat \lcdm model.  We have used them as an illustrative example in the absence of any parameter values obtained by constraining the present model with the observational data sets.

It can be said quite conclusively that this scalar field model resolves the initial condition problem, produces late-time acceleration with $\wphi = -1$ as predicted by the recent data as well as decreases rms mass fluctuation $\se$. This model is also successful in the context of the structure formation in the Universe. The model looks to be promising, but it has to be tested against observational datasets, and compared with \lcdm and other competing models in connection with the evidence criteria.

\providecommand{\href}[2]{#2}\begingroup\raggedright\endgroup

\end{document}